\documentclass[twocolumn,english,floatfix,prl,aps,showpacs]{revtex4}
\usepackage{graphicx}
\usepackage{amssymb}

\makeatletter

\usepackage{slashbox}
\usepackage{hhline}

\usepackage{babel}
\makeatother
\begin{document}

\title{Irrational mode locking in quasiperiodic systems}

\author{Creighton K. Thomas, A. Alan Middleton}

\affiliation{Department of Physics, Syracuse University, Syracuse, NY 13244, USA}

\begin{abstract}
A model for ac-driven systems, based on the Tang-Wiesenfeld-Bak-Coppersmith-Littlewood
automaton for an elastic medium, exhibits mode-locked steps with frequencies
that are irrational multiples of the drive frequency, when the pinning
is spatially quasiperiodic. Detailed numerical evidence is presented
for the large-system-size convergence of such a mode-locked step.
The irrational mode locking is stable to small thermal noise and weak
disorder. Continuous time models with irrational mode locking and
possible experimental realizations are discussed.
\end{abstract}


\maketitle
Extended non-equilibrium systems subject to spatially uniform drives
exhibit a vast range of spatiotemporal behaviors, including uniform
and static configurations, patterns periodic in space or time, and
a great variety of spatially and temporally chaotic phenomena \cite{CH-review,BPA}.
In the cases where the local fields have a periodic attractor under
a constant drive, the addition of an oscillating external drive of
period $\tau$ can lead to mode locking: the local fields are translated
by exactly $p$ periods over the time $q\tau$, with fixed $p,q\in\mathbb{Z}$,
for a finite range of external drive amplitudes. This rational $p/q$
mode-locked step is evident in the invariance of a dynamical variable
(a velocity or current) to changes in the strength of the external
driving force and is thus seen as a plateau in the average rate of
change of the fields when plotted versus a control parameter. Rational
mode locking has been seen in regular Josephson junction arrays driven
by microwave radiation (giant Shapiro steps) \cite{BRTL-giant_shapiro}
and a number of translation-invariant fluid and granular systems \cite{AT-granular}.
This mode locking can also be seen in systems with quenched disorder,
e.g., experiments on \cite{MRR-CDW_expt} and models of \cite{Coppersmith.Littlewood}
charge density waves (CDWs). A natural addition to these classes of
spatially uniform or periodic systems and spatially random systems
is the class of systems that have fixed quasiperiodic backgrounds.

Spatially quasiperiodic systems have been extensively explored in
a number of physical contexts. Quasiperiodic solids are of course
one example \cite{Quasicrystal.book}. The static and dynamic properties
of mesoscopic quasiperiodic systems have also been a subject of study.
In quasiperiodic wire arrays, the dependence of the normal-superconductor
transition temperature on external field has cusps at irrational densities
of the vortices \cite{nori,BBDLWC-wire_array}. Similar complex behavior
has been seen in pinning structures which consist of quasiperiodic
arrays of holes or magnetic (anti)dots added to a uniform superconductor
\cite{KGSPNSKK-antidots}. Quasiperiodic arrays of colloidal particles
can also now be constructed, using holographic optical traps \cite{Grier-quasiperiodic_colloids}.

We present here results on temporal mode locking in ac-driven
two-dimensional systems with quasiperiodic structure. In addition
to the typical rational mode locking, our simulations provide clear
evidence of mode-locked steps with $p/q$ approaching
an irrational value over a finite drive range in the large system limit.
These steps are robust to
small thermal noise and weak distributed disorder. It would be of
interest to search for these anomalous steps in experiments,
where an output frequency would be irrationally related to the input
frequency in the limit of large samples.

For speed of computation, we carried out our initial simulations using
a cellular automaton approximation of periodically driven overdamped
systems.  We first review its justification, based on arguments
developed by Tang, Wiesenfeld, Bak, Coppersmith, and Littlewood
(TWBCL) \cite{Coppersmith.Littlewood,TWBCL,BTW,coppersmith}.  The
automaton approximates a system with scalar degrees of freedom that
represent particle displacements or coarse-grained coordinates for an
elastic interface, such as the phase displacements in a charge density
wave, flux vortex displacements along a drive direction, or
fluid-fluid interfaces without overhangs \cite{Fisher}. The essential
features that are to be captured are displacements that can be
represented by an overdamped scalar field, local elastic interactions,
spatial variation (such as impurity disorder that screens elastic
interactions or a periodic background) that gives a natural length
scale, pinning by a potential that is periodic in the scalar field,
and a periodic external force. These general physical features
can be modeled by a
continuous-time driven Frenkel-Kontorova (FK) model
\cite{Floria-Mazo}, where $N$ dynamical variables $x_{i}$ (the
coarse-grained interface or particle displacement indexed by lattice
position $i$) evolve according to
\begin{eqnarray}
\dot{x_{i}} & = &
K\Delta^{2}x_{i}+h'\cos\left(kx_{i}-\beta_{i}\right)+F_{0}(t)\ ,\label{eq:FKEOM}
\end{eqnarray}
where $\Delta^{2}$ is the discrete Laplacian operator, $K$ is the
elastic constant, $h'$ is the magnitude of the pinning potential, $k$
determines the periodicity of the pinning, $\beta_{i}\in[0,2\pi)$ are
pinning phases, and $F_{0}(t)$ is a periodic driving force. Changing
variables by substituting $u_{i}=(kx_{i}-\beta_{i})/2\pi$,
$t\rightarrow2\pi Kt$, $h=h'/2\pi K$, and $F(t)=F_{0}(t)/2\pi K$
gives
\begin{eqnarray} \dot{u_{i}} & = & \Delta^{2}u_{i}+h\cos(2\pi
u_{i})+F(t)+b_{i}\ ,\label{eq:FKEOM2}
\end{eqnarray}
with
$b_{i}=(2\pi)^{-1}\Delta^{2}\beta_{i}$ in the range $(-z,z)$, where
$z$ is the number of nearest neighbors ($z=4$ for a square
lattice). Coppersmith \cite{coppersmith} has shown that in the limit
of large $h$ and $F(t)$ of the form
$F(t)=\sum_{k\in\mathbb{Z}}F\delta(t-k\tau)$ with ample relaxation
time $\tau$ between pulses, an automaton model is an appropriate
approximation to these driven FK dynamics. This claim is also
supported by numerical work, which shows very similar mode-locking
behavior for the continuous time model and the corresponding automaton
\cite{Coppersmith.Littlewood,MBLS}. In this limit, the dynamical
variable $u_{i}$ is near a minimum of the pinning potential,
$u_{i}\approx n_{i}\in\mathbb{Z}$, between drive pulses.  One then
defines an integer curvature variable $c_{i}$ by
\begin{eqnarray}
c_{i} & = & \Delta^{2}n_{i}+\left\lfloor b_{i}+F-h\right\rfloor
\ ,\label{eq:CurvatureMapping}
\end{eqnarray}
where $\left\lfloor
\right\rfloor $ is the floor function; this truncation of the net
force approximates the selection of local minima $n_{i}$ by peaks in
the pinning potential. The evolution of the $c_{i}$ between subsequent
pulses is the TWBCL automaton. At each time step, if the curvature
exceeds the critical value $z$, the variable $n_{i}$ advances (due to
the pulse and relaxation), redistributing the curvature ({}``grains of
sand'' \cite{BTW}) from a site $i$ to neighboring sites $j$.
Defining the toppling variable $U_{i}(t)=1$ at all $i$ where
$c_{i}(t)\ge z$ and $U_{i}(t)=0$ elsewhere, the update rule for
$c_{i}$ is
\begin{eqnarray}
c_{i}(t+1) & = &
c_{i}(t)+\Delta^{2}U_{i}\ .\label{eq:BTWDynamics}
\end{eqnarray}
In a
given step of this automaton, the activity or toppling rate,
corresponding to the velocity of the interface, is
$v_{t}=N^{-1}\sum_{i}U_{i}(t)$.  Given periodic boundary conditions
(BCs), the running dynamics of the system is fixed by the initial
conditions $\left\{ c_{i}(0)\right\} $, with $n_{i}=0$ in
Eq.~(\ref{eq:CurvatureMapping}) \cite{narayan.middleton}.

In the Bak-Tang-Wiesenfeld sandpile version \cite{BTW} (open BCs)
and in disordered driven
interfaces (reflecting BCs), the initial $c_{i}$ are
chosen from a Poissonian distribution. We choose the initial $c_{i}$
using Eq.~(\ref{eq:CurvatureMapping}) and quasiperiodic pinning
\begin{eqnarray}
b_{i} & = & 2z\textrm{ frac}\left(m\frac{P_{x}}{L_{x}}+n\frac{P_{y}}{L_{y}}\right)-z+\frac{1}{2N}\label{eq:OijDef}
\end{eqnarray}
on a 2D square lattice of size $N=L_{x}\times L_{y}$,
$i=(m,n)$, $0\le m<L_{x}$,
$0\le n<L_{y}$, and $\mathrm{frac}(x)=x-\left\lfloor x\right\rfloor $.
The latter two terms in Eq.~(\ref{eq:OijDef}) maintain $F\rightarrow-F$
symmetry. We choose the fractions $P_{x}/L_{x}$ and $P_{y}/L_{y}$
to be approximants to irrational numbers $\rho_{x}$ and $\rho_{y}$,
respectively. The approximants are found by truncating the continued
fraction representation
\begin{eqnarray}
\begin{array}{ccc}
\rho & \equiv & (r_{0},r_{1},r_{2},\ldots)\end{array} & \equiv &
r_{0}+\frac{1}{r_{1}+\frac{1}{r_{2}+\cdots}}\ .\label{eq:continued_fraction}
\end{eqnarray}
We focus here on the case where $\rho_{x}=\sqrt{2}-1=(0,2,2,2,\ldots)$
and $\rho_{y}=\sqrt{3}-1=(0,1,2,1,2,\ldots)$. 

\begin{figure}[tb]
\centering

\includegraphics[%
  width=3.25in,
  keepaspectratio]{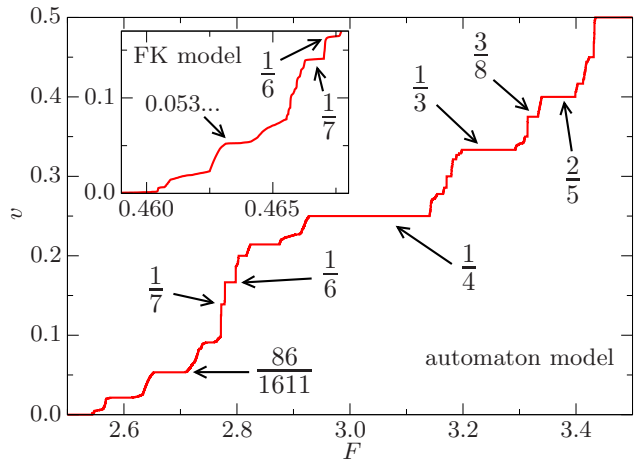}

\caption{\label{fig:QPStaircase}{[}color online{]} The activity staircase,
velocity $v$ vs.\ drive $F$, for the quasiperiodic automaton of
size $L_{x}\times L_{y}=169\times153$ with pinning given by the rational
approximants $(70/169,112/153)$ to $(\sqrt{2}-1,\sqrt{3}-1)$. Fractions
with small denominators generally have the widest mode-locked steps,
but the step at $86/1611$ is an anomalous step which does not fit
this pattern. Inset: Results for the FK-model,
Eq.~(\ref{eq:FKEOM2}), with $h=10$ and the same size and
quasiperiodic pinning; an anomalous
step with similar $v\approx0.053$ is seen.}
\end{figure}

Given pinning $\left\{ b_{i}(0)\right\} $ and drive $F$
that fix $c_i(t=0)$, one characterization of
the dynamics is the average activity rate $v=\langle v_{t}\rangle$
(toppling rate or interface velocity). For any finite system there
are only a finite number
of configurations, so there is at least one periodic attractor (all attractors
have the same $v$ \cite{AAM}). We repeat the
map, Eq.~(\ref{eq:BTWDynamics}), until a periodic orbit is found
and compute
$v(F) = M/(TN)$,
where $T$ is the period and $M=\sum_{t=1}^{T}\sum_{i}U_{i}(t)$
is the number of topplings in one period.
We find plateaus in $v$, i.e.\ mode-locked steps (Fig.~\ref{fig:QPStaircase}). Most of the wide steps seen are low-denominator
rationals, but there is at least one extra ``anomalous'' step which has a large
denominator for its width. 

\begin{figure}[tb]
\centering

\includegraphics[%
  width=3.1in,
  keepaspectratio]{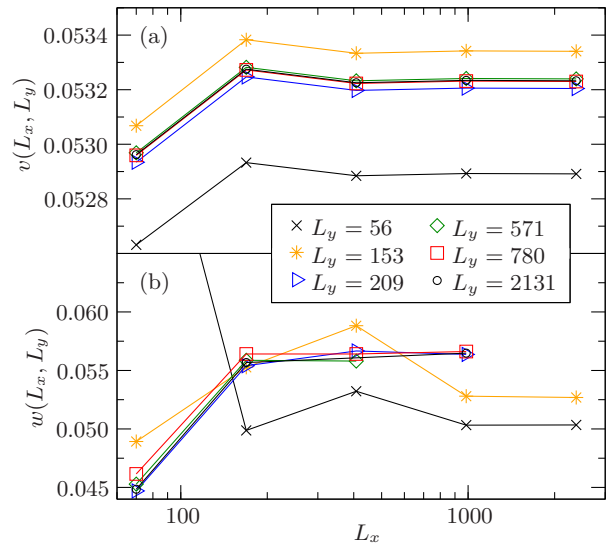}

\caption{\label{fig:scaling}{[}color online{]} The size dependence
of (a) activity $v(L_{x},L_{y})$, and (b) width, $w(L_{x},L_{y})$,
of the anomalous step. Each curve is for fixed $L_{y}$. The convergence
of $v$ and $w$ as a function of system size strongly
suggests that there is a well defined activity $\hat{v}$ and step
width $\hat{w}$ in the limit $(L_{x},L_{y})\rightarrow(\infty,\infty)$;
we conjecture that $\hat{v}\protect\notin\mathbb{Q}$, Eq.~(\ref{eq:IrratNum}),
and $\hat{w}\neq0$.}
\end{figure}

We find such a step, similar in position, size, and activity,
for all approximants to $(\rho_{x},\rho_{y})=(\sqrt{2}-1,\sqrt{3}-1)$
with $L_{x}\geq29$ and $L_{y}\geq56$. The activity $v(L_{x},L_{y})$
on this step is not fixed at a single rational (Table~\ref{cap:numbers});
the denominator and numerator for $v(L_{x},L_{y})$ grow with system
size, but $v$ appears to converge as $(L_{x},L_{y})\rightarrow(\infty,\infty)$
(Fig.~\ref{fig:scaling}). This strongly implies a limiting anomalous
step at an unusual activity rate.
The width of the selected
step also appears to converge (Fig.~\ref{fig:scaling}). Scanning
Table~\ref{cap:numbers} along the $\rho_{x}=\sqrt{2}-1$ approximant
direction (varying $L_{x}$, fixed $L_{y}$), it can be seen that
the numerators $p_{a,b}$ for $v(L_{x},L_{y})$ satisfy the recurrence
relation
\begin{eqnarray}
p_{a,b} & = & 2p_{a-1,b}+p_{a-2,b}\ ,\label{eq:Root2RecRel}
\end{eqnarray}
where $a$ ($b$) is the number of terms used in the approximation
of $\rho_{x}$ (respectively, $\rho_{y}$). The denominators $q_{a,b}$
obey the same relation. Similarly, along the $\rho_{y}=\sqrt{3}-1$
direction (varying $L_{y}$, fixed $L_{x}$) the recurrence relation
that agrees with all data is
\begin{eqnarray}
p_{a,b} & = & \left\{ \begin{array}{cc}
p_{a,b-1}+p_{a,b-2} & \textrm{if }b\textrm{ is odd}\\
2p_{a,b-1}+p_{a,b-2} & \textrm{if }b\textrm{ is even}\end{array}\right.\ ,\label{eq:Root3RecRel}
\end{eqnarray}
with the denominators $q_{a,b}$ also obeying the same relation.
Note that Eqs.~(\ref{eq:Root2RecRel},\ref{eq:Root3RecRel}) are
also the recurrence relations for the numerators and denominators
in, respectively, the approximants $P_{x}/L_{x}$ and $P_{y}/L_{y}$.
These recurrence relations suggest the temporal concatenation and
synchronization of subsystems, but the precise construction of such
spatiotemporal behavior is unclear to us.
Jointly solving Eqs.~(\ref{eq:Root2RecRel},\ref{eq:Root3RecRel})
using four velocities as seeds, we obtain the velocity $\hat{v}$
in the $(L_{x},L_{y})\rightarrow(\infty,\infty)$ limit
\begin{eqnarray}
\hat{v} & = & \frac{7903+379\sqrt{3}+142\sqrt{2}-181\sqrt{6}}{156238}\ ,\label{eq:IrratNum}
\end{eqnarray}
which is an algebraic irrational number, as our conjecture for the
mode-locking value in the limit of infinite system size for this choice
of $(\rho_{x},\rho_{y})$.

\begin{table}[tb]
\centering 

\begin{tabular}{|c||c|c|c|c|c|c|c|} \hhline{|-||-|-|-|-|-|-|-|} \large{\backslashbox{$\frac{P_x}{L_x}$\rule[-1.2ex]{0pt}{0pt}}{$\frac{P_y}{L_y}$\rule{0pt}{2.6ex}}}&  {\large $\frac{41}{56}$}&  {\large $\frac{112}{153}$\rule{0pt}{2.6ex}\rule[-1.2ex]{0pt}{0pt}}&  {\large $\frac{153}{209}$}&  {\large $\frac{418}{571}$}&  {\large $\frac{571}{780}$}& {\large $\frac{1560}{2131}$}& {\large $\frac{2131}{2911}$} \tabularnewline \hhline{:=::=:=:=:=:=:=:=:} {\large $\frac{12}{29}$}&  {\large $\frac{9}{167}$}&  {\large $\frac{22}{405}$\rule{0pt}{2.6ex}\rule[-1.2ex]{0pt}{0pt}}&  {\large $\frac{31}{572}$}&  {\large $\frac{84}{1549}$}&  {\large $\frac{115}{2121}$}& {\large $\frac{314}{5791}$}& {\large $\frac{429}{7912}$}\tabularnewline \hhline{|-||-|-|-|-|-|-|-|} {\large $\frac{29}{70}$}&  {\large $\frac{14}{266}$}&  {\large $\frac{32}{603}$\rule{0pt}{2.6ex}\rule[-1.2ex]{0pt}{0pt}}&  {\large $\frac{46}{869}$}&  {\large $\frac{124}{2341}$}&  {\large $\frac{170}{3210}$}& {\large $\frac{464}{8761}$}& {\large $\frac{634}{11971}$}\tabularnewline \hhline{|-||-|-|-|-|-|-|-|} {\large $\frac{70}{169}$}&  {\large $\frac{37}{699}$}&  {\large $\frac{86}{1611}$\rule{0pt}{2.6ex}\rule[-1.2ex]{0pt}{0pt}}&  {\large $\frac{123}{2310}$}&  {\large $\frac{332}{6231}$}&  {\large $\frac{455}{8541}$}& {\large $\frac{1242}{23313}$}& {\large $\frac{1697}{31854}$}\tabularnewline \hhline{|-||-|-|-|-|-|-|-|} {\large $\frac{169}{408}$}&  {\large $\frac{88}{1664}$}&  {\large $\frac{204}{3825}$\rule{0pt}{2.6ex}\rule[-1.2ex]{0pt}{0pt}}&  {\large $\frac{292}{5489}$}&  {\large $\frac{788}{14803}$}&  {\large $\frac{1080}{20292}$}& {\large $\frac{2948}{55387}$}& {\large $\frac{4028}{75679}$}\tabularnewline \hhline{|-||-|-|-|-|-|-|-|} {\large $\frac{408}{985}$}&  {\large $\frac{213}{4027}$}&  {\large $\frac{494}{9261}$\rule{0pt}{2.6ex}\rule[-1.2ex]{0pt}{0pt}}&  {\large $\frac{707}{13288}$}&  {\large $\frac{1908}{35837}$}&  {\large $\frac{2615}{49125}$}& {\large $\frac{7138}{134087}$}& \tabularnewline \hhline{|-||-|-|-|-|-|-|-|} {\large $\frac{985}{2378}$}&  {\large $\frac{514}{9718}$}&  {\large $\frac{1192}{22347}$\rule{0pt}{2.6ex}\rule[-1.2ex]{0pt}{0pt}}&  {\large $\frac{1706}{32065}$}&  {\large $\frac{4604}{86477}$}&  {\large $\frac{6310}{118542}$ }& {\large $\frac{17224}{323561}$}& \tabularnewline \hhline{|-||-|-|-|-|-|-|-|} {\large $\frac{2378}{5741}$}& {\large $\frac{1241}{23463}$}& {\large $\frac{2878}{53955}$\rule{0pt}{2.6ex}\rule[-1.2ex]{0pt}{0pt}}& {\large $\frac{4119}{77418}$}& & & & \tabularnewline \hhline{|-||-|-|-|-|-|-|-|} \end{tabular}

\caption{\label{cap:numbers}The system-size dependence of activity (velocity)
$v(L_{x},L_{y})$ at the anomalous step. The model is the TWBCL automaton
with quasiperiodic pinning. The fractions $P_{x}/L_{x}$ and $P_{y}/L_{y}$,
which determine the pinning, are approximants to $\sqrt{2}-1$ and
$\sqrt{3}-1$, respectively. To highlight the recurrence relation,
Eqs.~(\ref{eq:Root2RecRel}, \ref{eq:Root3RecRel}), the fractions
are not all written in lowest terms; in most cases, $\textrm{GCD}(p,q)=\textrm{GCD}(L_{x},L_{y})$.}
\end{table}

We have found
anomalous steps using other irrational pairs
$(\rho_x,\rho_y)$.
For small
system sizes, there are no steps that satisfy the recurrence
relations, but as the system size increases, irrational
candidates appear, with more than one candidate for a given
$(\rho_x,\rho_y)$.  It is even possible that there are an
infinite number of irrational steps with nonzero width in
the infinite size limit for typical irrational pinning
choices.
It would be very surprising if other
quasiperiodic pinning schemes (i.e., patterns for $b_i$) or
lattices did not also give irrational steps.
We have not found a way to predict, for a given
driving choice, the activity of any irrational mode-locked
step or the number of such steps.  One-dimensional automata
with nearest neighbor rules exhibit few rational steps and
in the simplest case should not exhibit irrational steps
\cite{MBLS}.

Numerically more challenging versions of this model include simulations
of the continuous-time FK model Eq.~(\ref{eq:FKEOM2}). We have
studied
the FK model using quasiperiodic pinning
variables identical to those for the automaton. We find results
similar to those for the automaton, with both rational and anomalous
steps converging to fixed width in a sequence of system sizes
(Fig.~\ref{fig:QPStaircase}, inset).

For applications to experiment, it is natural to ask about the robustness
of the irrational mode-locked step. To address this question we have
studied the effects of thermal noise and quenched disorder. We
model thermal noise by adding random topplings to the zero-temperature
parallel dynamics of the system, Eq.~(\ref{eq:BTWDynamics}). These
random topplings occur at finite rate $\eta$ per time step at each
site. This noise randomly increases or decreases
the displacement variable $n_{i}$ by one period. To
implement this, we add backwards topplings,
which take place when $c_{i}\leq-z$. 
The toppling variable $U_{i}(t)$ used in Eq.~(\ref{eq:BTWDynamics}) is
modified:
in addition to setting $U_{i}(t)=1$ at all $i$ where $c_{i}(t)\ge z$,
one sets $U_{i}(t)=-1$ at all $i$ where $c_{i}(t)\leq-z$, with
$U_{i}(t)=0$ for $-z<c_{i}(t)<z$.

\begin{figure}[tb]
\centering

\includegraphics[%
  width=3.25in,
  keepaspectratio]{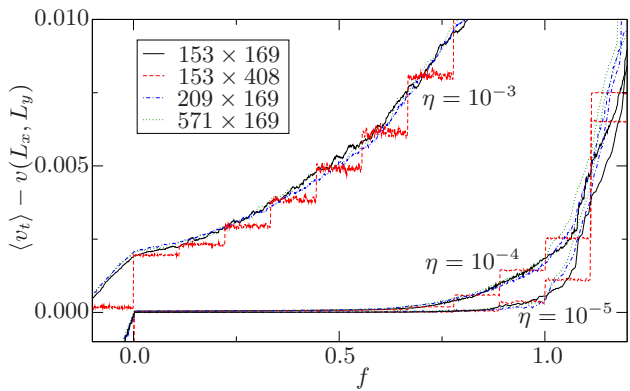}

\caption{
\label{fig:therm_noise}{[}color online{]} The effect of thermal noise
on the anomalous step for dimensionless noise values $\eta = 10^{-5}$,
$10^{-4}$, $10^{-3}$ for
 system sizes defined in the key. The horizontal axis is the scaled
force $f$ which is linear in $F$ with each zero-temperature step starting at
$f=0$ and ending at $f=1$. For easier comparison, the
zero-temperature activity $v(L_{x},L_{y})$ has been subtracted.  For each value
of noise, there is similar rounding of the plateau for all system sizes.
}
\end{figure}

Automaton models with period $q>2$ are unstable to thermal noise
in general \cite{BGHJM}, but we find that for small thermal noise
the change in the velocity is limited.
Weak thermal noise, which leads to
nucleation of out-of-phase regions \cite{BGHJM} and subsequent
motion of domain walls, destroys exact periodicity, but
the anomalous steps
can still be clearly identified
(Fig.~\ref{fig:therm_noise}).
Numerical simulations and analysis of similar systems, such as depinning
in CDWs with weak noise, also show small corrections to $\langle v\rangle$
on most of the step and rounding near small-denominator step edges
\cite{therm_stability}.

The effects of quenched disorder were included by modifying the pinning
variables $b_{i}$, to model jitter in both pinning phases $\beta_{i}$
and coupling $K$. Weak distributed disorder does not affect the activity
$v(L_{x},L_{y})$ significantly, with 97\% of the step unchanged for
0.1\% Gaussian noise in the \textbf{$b_{i}$} with $(L_{x},L_{y})=(169,153)$.
However, irrational locking is sensitive to strong rare disorder:
we find that randomizing the $b_{i}$ at even one site in large systems
drastically reduces the width of the anomalous step. This is to be
contrasted with Poissonian initial conditions, which are maximally
random, and thus have steps that are not sensitive to strong disorder,
though wide large-denominator steps are not present.

Given the general applicability of the FK model and the generic features
of the automaton approximation, our results suggest that irrational
mode locking might be seen in a number of experimental systems. The
most likely candidates are constructed mesoscopic systems, where thermal
fluctuations are small enough and there is precise experimental control.
As noted, quasiperiodic superconducting wire arrays and Josephson-junction
arrays exhibit interesting static features as a function of applied
field; dynamic ac transport experiments might well be modeled by elastically
or plastically interacting vortices \cite{ROH,RR} that would realize
the conditions necessary for irrational mode locking. Recently \cite{Grier-quasiperiodic_colloids},
quasiperiodic colloidal particle arrays have been constructed using
holographic optical traps. Experiments for dc driven particles in
periodic arrays \cite{Grier2} suggest that experiments in
a quasiperiodic background may be used to search for irrational mode locking
in a collective system.

We have studied a broadly applicable model of dynamics in an extended
system, where we have chosen a quasiperiodic spatial background. We
find anomalous mode-locked steps that appear to converge to an irrational
frequency relative to the drive frequency, in the limit of large system
sizes. This step is stable to modest amounts of both thermal noise
and quenched disorder.
One avenue for comparison with experiment would be to study automaton models on
quasiperiodic tilings (see \cite{Joseph} for the $v\rightarrow 0$ limit).
There are a number of experimental systems
that are closely related to this model and in which such irrational
mode locking might be seen.

We thank Ofer Biham for suggestions. This work was supported in part
by NSF via DMR-0219292, 0606424.

\end{document}